# Toward a Rational and Ethical Sociotechnical System of Autonomous Vehicles: A Novel Application of Multi-Criteria Decision Analysis

Veljko Dubljevic[1], George F. List[1], Jovan Milojevich[2], Nirav Ajmeri[3], William Bauer[1], Munindar P. Singh[1], Eleni Bardaka[1], Thomas Birkland[1], Charles Edwards[4], Roger Mayer[1], Ioan Muntean[5], Thomas Powers[6], Hesham Rakha[7], Vance Ricks[8] and M. Shoaib Samandar[1]

[1]North Carolina State University
[2]Oklahoma State University
[3]University of Bristol
[4]University of North Carolina at Chapel Hill
[5]University of North Carolina at Asheville
[6]University of Delaware
[7]Virginia Tech
[8]Guilford College

## 1. Introduction: MCDA and the problems of AVs

The expansion of artificial intelligence (AI) and autonomous systems has shown the potential to generate enormous social good while also raising serious ethical and safety concerns [1-4]. AI technology is increasingly adopted in transportation. Kamalanathsharma *et al.* [5] conducted a survey of various in-vehicle technologies and found that approximately 64% of the respondents used a smartphone application to assist with their travel. The top-used applications were navigation and real-time traffic information systems. Among those who used smartphones during their commutes, the top-used applications were navigation and entertainment.

There is a pressing need to address relevant social concerns to allow for the development of systems of intelligent agents that are informed and cognizant of ethical standards. Doing so will facilitate the responsible integration of these systems in society. To this end, we have applied Multi-Criteria Decision Analysis (MCDA) to develop a formal Multi-Attribute Impact Assessment (MAIA) questionnaire for examining the social and ethical issues associated with the uptake of AI. We have focused on the domain of autonomous vehicles (AVs) because of their imminent expansion [6]. However, AVs could serve as a stand-in for any domain where intelligent, autonomous agents interact with humans, either on an individual level (e.g., pedestrians, passengers) or a societal level [7].

MCDA has been proposed as a method to study the assessment of harms and risks [8-10; 11,12]. The basic tenet is that MCDA, along with qualitative techniques, can provide defensible insights about the way people see the multi-faceted impacts of technological change. Many improvements have already been made since the papers describing this method were initially published. For instance, studies have 1) expanded the criteria [11], 2) included relative importance (or weights) of different harms [9], 3) made comparisons of the harm/benefit ratios [12], and 4) reported the need to include perceptions of all relevant stakeholders [13].

One of the strengths of MCDA is that it can systematize a process that encompasses large areas of knowledge in a transparent manner, allowing for replication and improvement of the methodology [13]. MCDA breaks down complex evaluations into a series of smaller, more easily assessed issues, thus enhancing the reliability and validity of the results.

By utilizing both qualitative and quantitative analyses, we have expanded the utility of MCDA, giving it the potential to drastically improve the ethical evaluations of transformative change, illustrated here in the context of AV technology. The MAIA questionnaire provides an evidence base for impacts, including harm-over-benefit ratios. Notably, it addresses the drawbacks identified in the literature critical of the MCDA methodology, such as lack of attention to situational factors [*14*], value judgments [*15*], and additional stakeholders [*16-18*].

There is a substantial need to apply the kind of approach embodied by MAIA to AVs, for the following reasons. As AVs are implemented in various types of transportation systems, the degree of direct interaction with humans (e.g., pedestrians) and human operated vehicles (connected non-autonomous vehicles (CVs) and traditional vehicles) grows in complexity both intrinsically and due to the combinatorial complexity introduced by large numbers of vehicles. Therefore, controlling the behavior of AVs becomes inherently more complex, and the potential for harm to humans increases. In simpler versions of the transport system (e.g., robotic single-lane freeways), it is possible to consider the devices to be automatic, precisely carrying out the instructions of the owners, according to relatively simple programming. However, in complex urban environments, where the interactions are far more complex, humans frequently take actions outside the rule set to resolve conflicts. Therefore, successfully implementing AVs requires accommodating unpredictable situations that may occur as a result of human behavior and decision-making.

AVs will transform the lives of many people [*19,20*]. Although they have the potential to save many lives, they also raise important new safety and ethical dilemmas. Due to the inherent human factors involved, the successful implementation of AVs is not only an engineering issue but a social, political, and ethical issue as well. The perspectives of multiple disciplines are required to craft detailed assessments of relative impacts for implementation of AVs in different types of controlled and uncontrolled transportation environments. Understanding the societal and ethical implications of AVs (and any AI system) inherently involves many distinct issues: the nature and capabilities of these technologies (computer science, engineering), how humans can and should use them (ethics), how humans will behave in response to the presence of AVs in the traffic stream (social sciences), and the technology's impact on socio-economic structures (political science, economics). Thus, producing new and relevant knowledge in this area requires the expertise originating in multiple disciplines [*21*].

Expert opinions are periodically obtained on emerging technologies to provide valuable insights [*22-24*]. However, a comprehensive methodology for comparing heterogeneous harms and benefits relative to different stakeholders has been lacking. Previously, expert assessments of AV technology have been based on fictional future scenarios so that relevant policies could be discussed [*25*], as opposed to identifying how policies adopted in the present could shape the future, or how each policy option compares to the current status quo in terms of relevant criteria (see Table 1).

To fill this gap, we elicited expert opinions about the impacts of AVs through a Delphi exercise and consensus workshop, resulting in operational evidence regarding the moral, social, and economic benefits and harms of AVs. The identification of relevant facts and values—the task for which disciplinary experts are essential—helps guide complex evaluations, reduces confounds and biases, and clarifies uncertainties [*26*].

Our assumption, at least for the foreseeable future, is that it is unrealistic to expect AVs to completely replace traditional non-autonomous motor vehicles. We expect that AVs will operate in a heterogeneous environment alongside traditional vehicles, as well as cyclists and pedestrians. Traditional vehicle technology is assumed to be robust and wanted not merely for economic benefits but also for psychological reasons, such as the 'joy of driving' [27]. Thus, we agree with Samandar and colleagues that "a mixed traffic fleet is likely to be the predominant scenario for the foreseeable future." [28]

Studies focused on other countries have generated assessments that are interesting but not necessarily applicable to the U.S. context. For instance, the German Federal Ministry of Transport and Digital Infrastructure appointed a national ethics committee for automated and connected driving to develop and issue a code of ethics. This code states that "protection of individuals takes precedence over all utilitarian considerations" and "automated driving is justifiable only to the extent to which conceivable attacks, in particular manipulation of the IT system or innate system weaknesses, do not result in such harm as to lastingly shatter people's confidence in road transport." [29] Such guidance is interesting, yet there is no mention of how it is to be implemented, raising concerns of its feasibility. Moreover, it fails to address important issues such as how AV technology could be programmed to resist malicious actors, such as terrorists [30,31], or how social justice issues can be safeguarded during the introduction of AVs into the socioeconomic system [32]. The European Union [24] and Australia [33] have also developed expert-assessed scenarios intended to guide policy makers in regulating AV technology. Groups of experts are very important, but they are better used in assessing the importance of harms and benefits, as we have done.

## 2. Methodology: The Multi-Attribute Impact Assessment

We developed a novel application of the MCDA method, which we call the Multi-Attribute Impact Assessment (MAIA) questionnaire, to assess the impacts of AV technology. We identified 21 impacts for which we sought expert opinions about their importance (See Supplementary Table ST1). We followed an iterative process that began with the first author of this paper preparing an initial list of harms and benefits based on the AV ethics literature and relevant agency reports [ 34, 35]. The list was discussed at length by a sample of six experts (first six authors of this paper), then revised based on the feedback. It was subsequently piloted in a Delphi survey with the full panel of 19 experts, again revised based on feedback and discussed at length during the consensus workshop (see below). The final list of impacts is categorized into 13 harms and eight benefits, as shown in Table I.

**Table I: The harms and benefits assessed**

| Q1 | Harms of vehicle related mortality (e.g. driver or passenger deaths on the road) |
|----|----------------------------------------------------------------------------------|
| Q2 | Harms of vehicle specific damage (e.g., costs of damage to property) |
| Q3 | Harms of vehicle related damage (e.g., damage to natural environment) |
| Q4 | Harms of vehicle system encroachment on human living (e.g, reduction of urban walkability) |
| Q5 | Harms of vehicle related occupational injuries (e.g., sedentary lifestyle of drivers) |
| Q6 | Harms of vehicle related lack of status (e.g., elderly losing driver's licenses due to visual impairments) |
| Q7 | Harms of vehicle related loss of time or productivity (e.g, time spent in traffic jams) |
| Q8 | Harms of vehicle related loss of social engagement (e.g., time spent isolated from |

| | others) |
|-----|---------|
| Q9 | Harms of vehicle related injury to others (e.g., hit and run incidents) |
| Q10 | Harms of vehicle related economic costs (e.g., maintenance costs) |
| Q11 | Harms of vehicle related changes to community (e.g., marginalization of specific communities) |
| Q12 | Harms of vehicle related crime opportunities (e.g., sexual assault by ride-hailing service drivers or passengers) |
| Q13 | Harms of vehicle related economic changes (e.g., loss of jobs by drivers) |
| Q14 | Benefits of promoting societal value (e.g., increase in economic activity) |
| Q15 | Benefits of minimizing negative societal impacts (e.g., decrease in pedestrian injury and death) |
| Q16 | Protecting the interests of users (e.g., drivers) |
| Q17 | Advancing the preservation of the environment (e.g., reducing traffic jams) |
| Q18 | Maximizing the progress of science and technology (e.g., increasing data quality) |
| Q19 | Engaging relevant communities (e.g., pedestrians, business communities) |
| Q20 | Ensuring oversight and accountability (e.g., preventing or limiting irresponsible uses) |
| Q21 | Recognizing appropriate governmental and policy roles (e.g., bringing public attention to transportation issues) |

Concurrently, we explored four operational scenarios or regulatory environments that might be implemented during the deployment of AVs. They are described in Table 2.

**Table II: Operational scenarios and regulatory environments explored**

| # | Definition | Description |
|---|-----------|-------------|
| 1 | Status Quo (S-Q) | The transportation system as it is currently, with non-AVs. |
| 2 | Unfettered AVs (U-F) | A transportation system in which there is no regulation and so implementation is unfettered and left to commercial entities (i.e., the tech industry). |
| 3 | Regulated privately owned AVs (R-P) | A transportation system which is regulated so that AVs are owned much like traditional passenger vehicles. They must be inspected and there are only certain "areas" where they can be operated. |
| 4 | Regulated fleet owned AVs (R-F) | A transportation system which is regulated so that AVs are owned only by commercial fleets, with stringent inspections, and there are designated areas where they can be operated. |

Note: In scenarios 2-4, we assume that traditional non-autonomous vehicles continue to operate in addition to AVs.

These categories were based on our sense of how AV technology is likely to be introduced. Beyond the *status quo* (scenario 1), the first AV condition (scenario 2) assumes no regulatory control will be exercised and commercial entities will "push" the development and deployment. Implicitly, anyone (any entity) would be able to purchase and operate such vehicles anywhere.

The second and third AV conditions (scenarios 3&4) assume that regulatory control will be applied and either individuals (scenario 3) or only commercial operators (scenario 4) can own the AVs. Scenarios 3 and 4 assume SAE level 4, meaning that the vehicles can operate on a portion of the highway network [35]. In scenario 3, companies can also own them, like car rental and ride sharing companies, but there is no prohibition against people having AVs. In scenario 4 only commercial operators can own AVs; no personal ownership is allowed. The categories are silent insofar as the level of market penetration is concerned; but implicitly, the vehicle population has enough AVs such that their operational impact is visible. We elected not to include SAE Level 5, full autonomy, as one of the scenarios because it seems far off in the future compared with the status quo.

## 3.  Results

A consensus emerged that certain forms of AV implementation would be less harmful than others. Namely, the regulated private or fleet owned policies (scenario 3 or 4) would be better than either the current transportation system (scenario 1) or the haphazard or unfettered AV implementation (scenario 2). The stacked histograms in Fig. S1, which show the harms of different AV technology implementation measured on a 4-point scale, summarize this finding.

Similarly, the regulated, fleet owned scenario (scenario 4) would produce the greatest benefits (see Fig. S2). A follow-up survey that used a 10-point scale produced similar results. See Fig. S3.

The harm and benefit assessments were open ended. Respondents were allowed to scale their total assessments on any basis. To make this clear, whereas one respondent could have used "1" as the maximum for each, another could have used "100". With 21 criteria, this means the first respondent would have a total ranging up to 21; the second, up to 2100. We wanted to see if the respondents were similar in their relative assessments of the importance of the 21 criteria. Hence, we scaled the assessments to a total of 100 for each. Fig. S4 shows the "weight profiles" that emerged. For each respondent, the profile shows the percentage distribution of importance among the 21 criteria for a given respondent. If one of the respondents had made the assessments all equal, the "profile" would be a straight line. A "quicker rise" for a given criterion implies that it has more importance, a slower rise, less importance. The main conclusion we draw from this figure is that, except for a couple of respondents, they all had a similar sense of the relative importance of the impacts. Respondent 5 (medium blue and the highest) gave the greatest aggregate importance to the harms (highest total percent by impact 13). Respondent 14 (dark orange, and the lowest) gave the greatest importance to the benefits (lowest total percent by impact 13).

### 3.1.    Harms

Two "harm" assessment surveys were administered. The harms were impacts 1-13. One survey used a 4-point scale (0-3) for each impact where zero was "no harm" and 3 was "extreme harm" (Cronbach's alpha, $\alpha = 0.863$). The other used a 10-point scale where 1 was "no harm" and 10 was "extreme harm" (Cronbach's alpha, $\alpha = 0.916$). The assessments were done sequentially, with the 4-point scale having been assessed first. Since the findings from the 4-point scale were shared with the participants before the 10-point survey was administered, there could be impacts from shared feedback; as demonstrated by the results of the Cronbach's alpha reliability tests.

We analyzed the harm responses in several ways. The first was in terms of the relative importance of the harms; another was by "scenario." The description of the harms is presented as in Table 1 (questions 1-13), and the four scenarios are in Table 2.

The mean values and standard deviations based on the 4-point scale (0-3) are shown in Fig. S5. A higher score means greater harm. The harm with the greatest reduction due to AVs is 6, lack of status loss (e.g., elderly losing driver's licenses due to visual impairments). This makes sense; AVs provide a significant boost in mobility for these people. The one where the impacts are mixed or minimal is 11, harms related to changes to community (e.g., marginalization of specific communities). The respondents saw no clear trend in this impact. The impact with the greatest variation in impact assessment was 3, vehicle damage (e.g., to the natural environment), we suppose this is because of differences in perceptions about the technology and how it will be used. Harm 13 stands out as having characteristics different from the others. It pertains to harms of vehicle related economic changes (e.g., loss of jobs by drivers). Hence, it is not surprising that its impacts are different. The aggregate assessment of differences among scenarios will be addressed later, but it seems clear that scenario 4, which involves a regulated commercially owned fleet, has the greatest reduction in harms.

Fig. S6 shows the same information but on a 10-point scale. The 1-10 results were remapped to 0-9 so that the low end of both assessments was 0. Strikingly different are the assessments for criterion 1 (more spread) and 3 (a higher sense of harm for the status quo). Otherwise, the pattern is similar. Moreover, as before Scenario 1 has the greatest harms, followed by Scenarios 2, 3, and 4, roughly in that order.

For a broader brush, we computed the sums by respondent for all the harms (the sum of the responses to questions 1-13). A maximum of 52 (13*4) was possible; and a minimum of 0. We then computed the average of these values and the standard deviation. Fig. S7 shows the results for both the four-point scale (0-3) and the ten-point scale (0-9). The trends in the average among the four scenarios is the same in both cases. The greatest harms are associated with the status quo (scenario 1); the least with the regulated / fleet owned scenario (scenario 4). These findings are consistent with visual inspections of Fig. S4 and S5. One noticeable difference is that the spread between the scenarios is larger in the 10-point case than in the 4-point instance. The trends in the standard deviations are also similar except that, in the case of the ten-point scale, the standard deviation for the laissez faire option or unfettered AVs (scenario 2) is higher than it is for the other three scenarios, whereas in the four-point assessment it is similar to the others. This could be an impact of the shared feedback on the 4-point survey.

## 3.2.  Benefits

Two "benefit" surveys were administered. As with the harms, one was on a 4-point scale (0-3) where zero was "no benefit" and 3 was "drastic benefits" (Cronbach's alpha, $\alpha = 0.901$); the other was on a 10-point scale where 1 was "no benefit" and 10 was "drastic benefits" (Cronbach's alpha, $\alpha = 0.941$).  For purposes of the presentation here, the 10-point scale is re-scaled to 0-9 so that "0" is common between the two surveys. For the "status quo" scenario, the benefits were not assessed as it was assumed that the status quo is the baseline.

Fig. S8 shows the benefit assessments based on both the 4-point and 10-point scales. Benefit 1 maps to impact or question 14 and benefit 8 to impact or question 21 listed in Table 1. We find that the greatest benefits are associated with 4 and 7, i.e., advancing the preservation of the environment (e.g., reducing traffic jams) and ensuring oversight and accountability (e.g., preventing or limiting irresponsible uses), respectively. These benefits are greater for the regulated scenarios (3 and 4) than for the unregulated one (2). Moreover, in a broader sense, the benefits for scenario 4 (regulated / fleet owned) are the greatest followed by scenario 3 (regulated / privately owned) and then scenario 2 (laissez faire or unfettered). The one benefit where scenario 2 produces comparable or higher benefits is 1, promoting societal value (e.g., increase in economic activity). Intuitively, respondents perceived that deregulated development would produce the most innovation and capital investment.

### 3.3.    Overall assessment

The summative question is: what does the survey suggest is the "best" scenario weighing the harms and benefits? There are many ways to answer that question [*36*]. One possible approach is to take the harm and benefit value assessments, by respondent, and combine them with the corresponding weights (by respondent), then we obtain sums of the results for the harms and the benefits. Admittedly, this is "problematic" in that the weights for the harms and benefits were assessed together; and here, they have been normalized to sum to one. But that may not be "bad" or "wrong." It can be argued that forcing them to sum to 1 provides, implicitly, the respondent's sense of the relative value of the eight benefits versus the 13 harms. Further surveying will reveal valuable information about this issue.

In this instance the "total of the weighted harms" has been plotted against the "total of the weighted benefits" for the four scenarios, based on the "weighted value assessments" of the respondents.  Fig. 1 plots the sum of these "weighted value assessments" for the harms against the "weighted value assessments" for the benefits. The message seems clear. The regulated-fleet owned scenario (4) is perceived to have greater benefits and lesser harms among all four options. It is slightly better than the regulated-personally owned scenario (3) and clearly better than the *laissez-faire* or unfettered scenario (2), especially insofar as the harms are concerned. (Of course, the status quo scenario has no benefits, and its harms are perceived to be the largest, significantly so in the case of the 10-point based assessment).

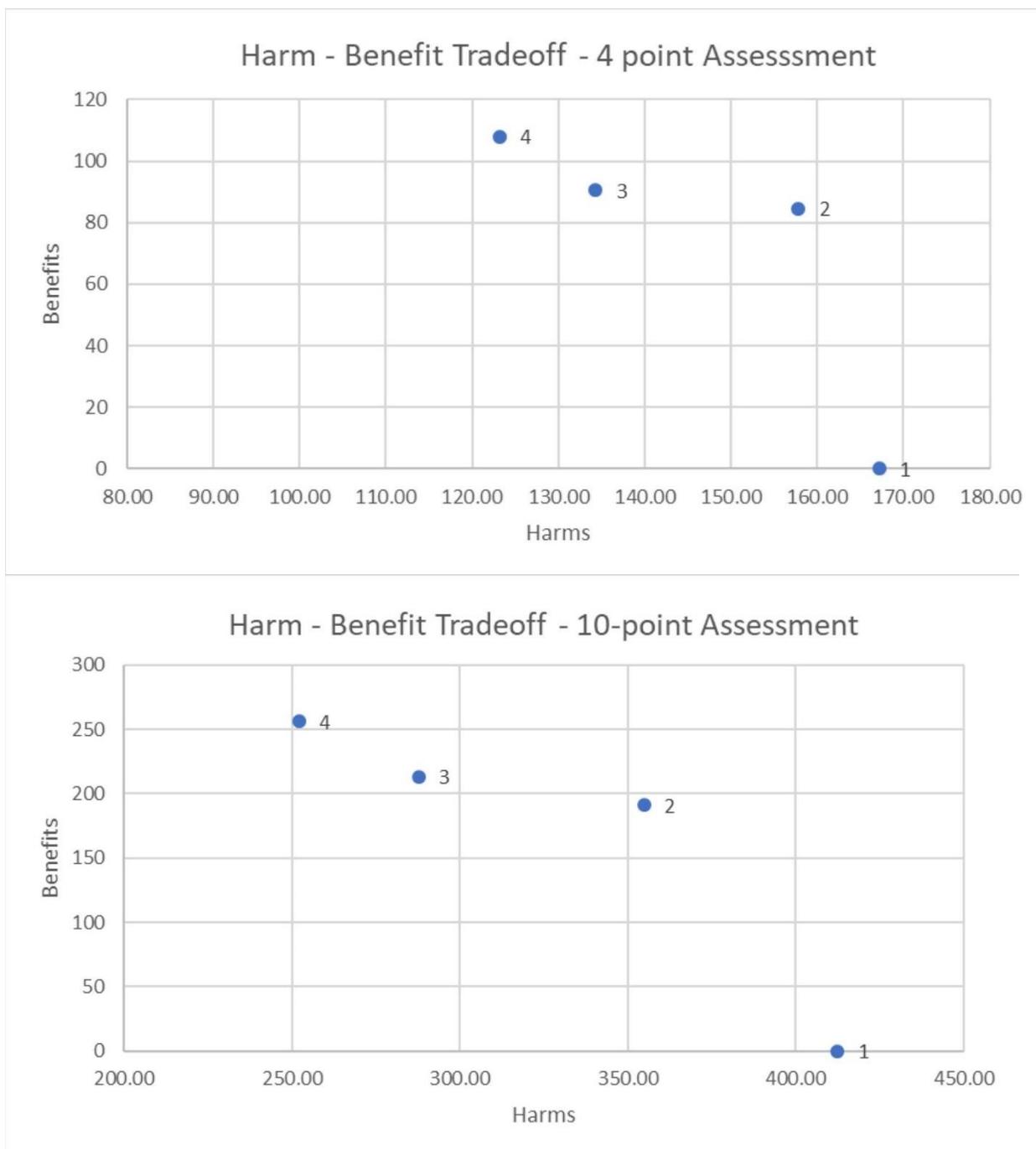

Figure 1: Harm/Benefit Tradeoffs for both the 4-point and 10-point assessments

## 4. Discussion and concluding remarks

Through the introduction of AVs, society is exposing itself to an array of risks. Despite the excitement surrounding this technology, there are many unanswered questions about whether it will be both beneficial and safe. Even though there are expectations of overall benefits to society from the deployment of AVs, some groups compared to others could experience (much) higher costs relative to benefits [*37*]. These negatively affected groups will include those whose livelihood depends on traditional motor vehicles. In particular, a significant number of drivers (and, by extension, family members depending on their economic activity) will be affected by the introduction of AVs. There are approximately 1.7 million truck drivers in the United States [*38*], with about 800,000 involved in truck transportation. A potentially exacerbating factor is that there is currently a shortage of truck drivers [*39*]. This could drive motivation to get AVs on the road en masse promptly, thus pushing current professional drivers out sooner. Although AVs will create new jobs in the trucking industry and other industries [*22*], it is an open question if these new jobs would outnumber those lost due to AVs; indeed, that seems unlikely. However, many driving jobs are perceived to be unsatisfying and potentially unhealthy (e.g., due to high incidence of sleep apnea and obesity), making their eradication an overall positive outcome if other employment opportunities are available [*33*].

Positive or negative utilities [*40*] related to the changes wrought to society needed to be assessed in a free and open discussion by a multi-disciplinary panel of experts. Our results point to the need to better address how the public views the trade-offs between (1) safety; (2) physical ecology (environmental issues); (3) social ecology; (4) economic issues; and finally (5) the specific impacts for groups that will be most affected by AV implementation (e.g., professional drivers).

We contend that it is essential to increase the public's confidence that the values of a pluralistic society are accounted for in the development of AV policies. This can be accomplished by 1) bringing society into the identification of norms surrounding AVs [*41, 42*] and 2) accounting for multiple elements of moral decision-making [*43*]. Regarding point 1, although expert groups like the one we assembled do not nearly represent society as a whole, if such a group is large enough and selected carefully it does represent an important slice of society that policymakers should pay attention to. Regarding point 2, such an expert group brings diverse, refined perspectives to moral decision-making that can only increase the reliability of the assessment by ensuring the most important considerations and values are brought to the surface.

Several states in the U.S. have started the process of legislating AVs, most notably designating the manufacturer of a vehicle operated by an automated driving system as the vehicle's sole driver, and limiting this special legal framework to motor vehicle manufacturers that deploy their vehicles as part of fleets within specific geographic areas [*44*]. Our work provides valuable data that should inform policy makers of concerns and potential benefits of AV technology in specific implementation strategies, and this could improve the quality of the democratic policymaking process. We recommend that state legislatures and the federal government strongly consider incorporating our results regarding technology development scenarios as well as the MAIA questionnaire into their deliberations about the impact of AVs.

# Appendix: Supplementary Information

## Supplementary methods section
We engaged 19 leading researchers from diverse backgrounds (in terms of discipline, gender and ethnicity) to participate in a consensus-building workshop on the NC State campus on 21 Feb 2020. We selected this many because 19 is near the upper limit espoused by Phillips [*45*] for effectiveness in expert-based decision analysis studies. The experts discussed the criteria and the scenarios at length during the workshop. Five surveys were administered in conjunction with the workshop: 1) weights among the criteria, 2) a 4-point assessment of harms, 3) a 4-point assessment of benefits, 4) a 10-point assessment of harms, and 5) a 10-point assessment of benefits. Subsequent to the workshop, an additional survey of weights limited to 100% total for all criteria was conducted.

The Delphi method was used to generate the responses. The participants were briefed on the results of the $n^{th}$ survey before the $n+1^{st}$ survey was administered. Their responses were converted to a 100-point scale, the participants were briefed about the result, and the survey was repeated with a second and a third wave. The repetition of rankings (using a 4-point scale and a 10-point scale) helps reduce potential biases in the impact assessments.

## Acknowledgements
This study was funded by North Carolina State University through Research and Innovation Seed Funding and the Kenan Institute for Engineering, Technology & Science.
The authors thank Abby Scheper, Abigail Presley, Leila Ouchchy, Joshua Myers, and Elizabeth Eskander for research assistance. Additional thanks to Missy Cummings, Stephanie Sudano, Joseph Hummer and Michael Struett for their valuable input during the workshop. Special thanks to the members of the Neuro-Computational Ethics research group for their feedback on an earlier version of the paper.

Table ST1: Composition of the Expert Pool

| Field of expertise | Number |
|---|---|
| Political science | 4 |
| Civil/transportation engineering | 7 |
| Philosophy/Ethics | 5 |
| Computer Science/AI | 3 |
| Organizational Behavior | 1 |

Note: The expert pool was diverse in the terms of ethnicity, gender, and other demographic variables

Figure S1: Harms of different AV technology implementation, 4-point scale

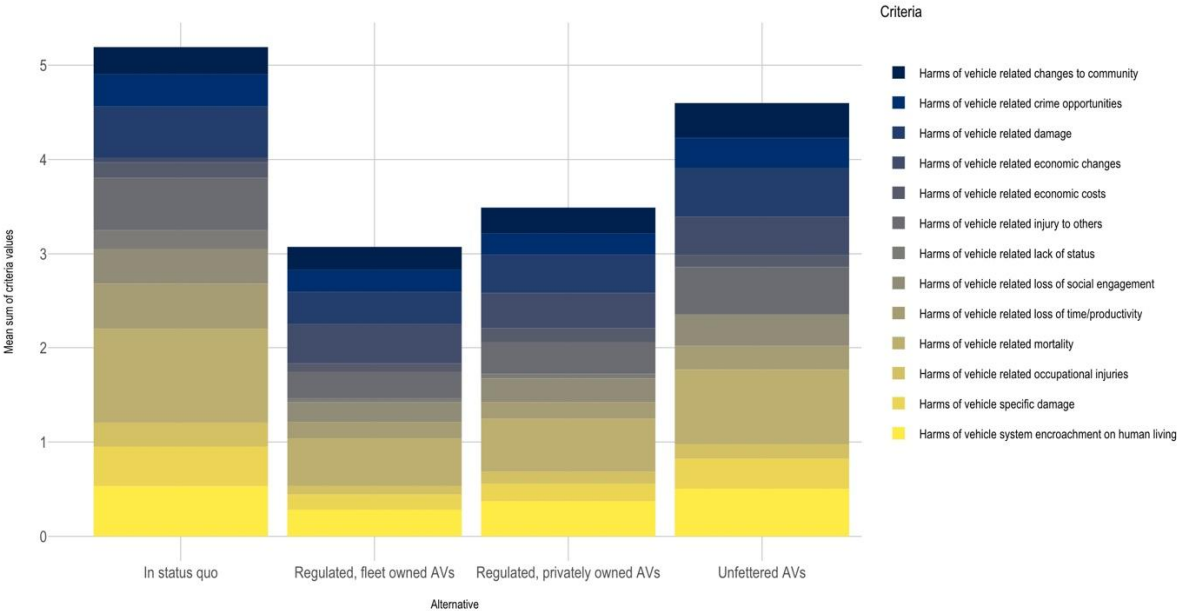

Figure S2: Benefits of different AV technology implementation, 4-point scale

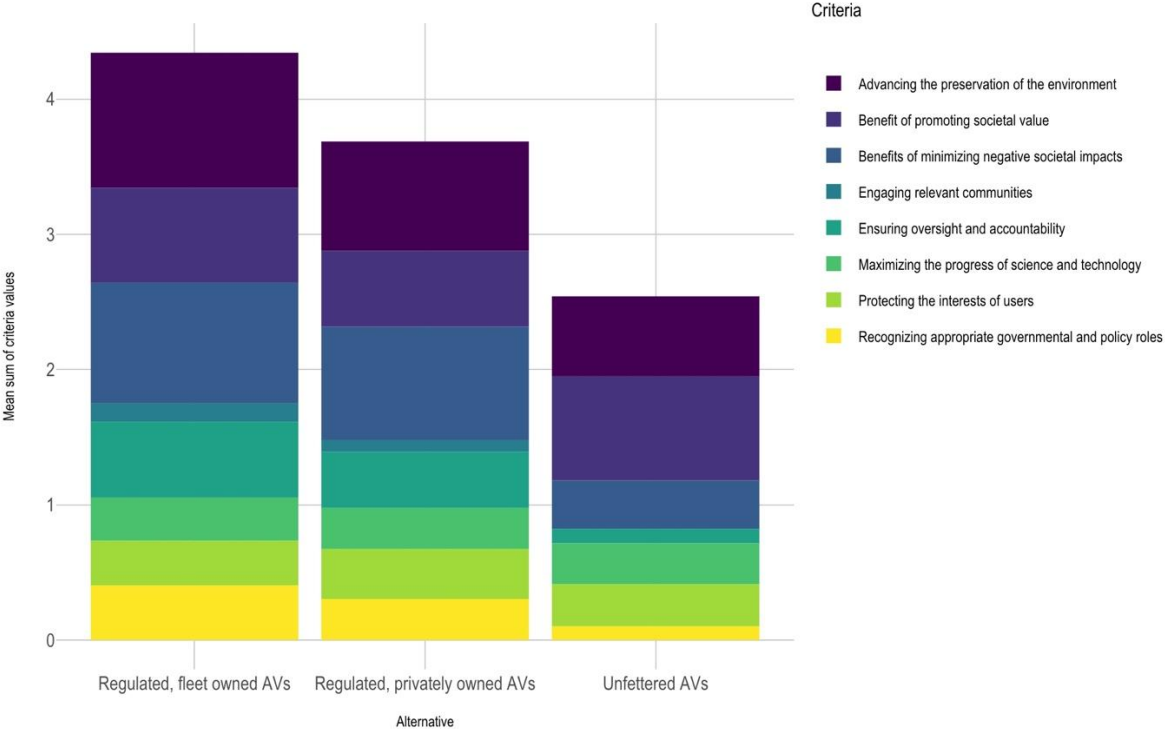

Figure S3 Harms and Benefits in a repeated, 10-point scale

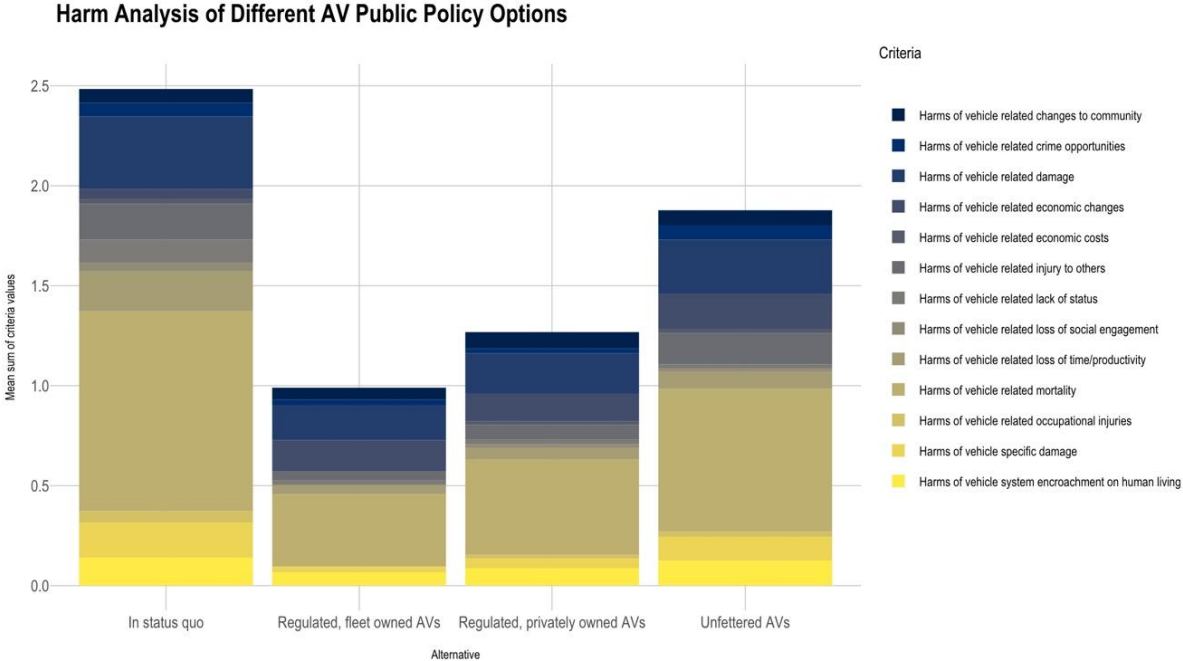

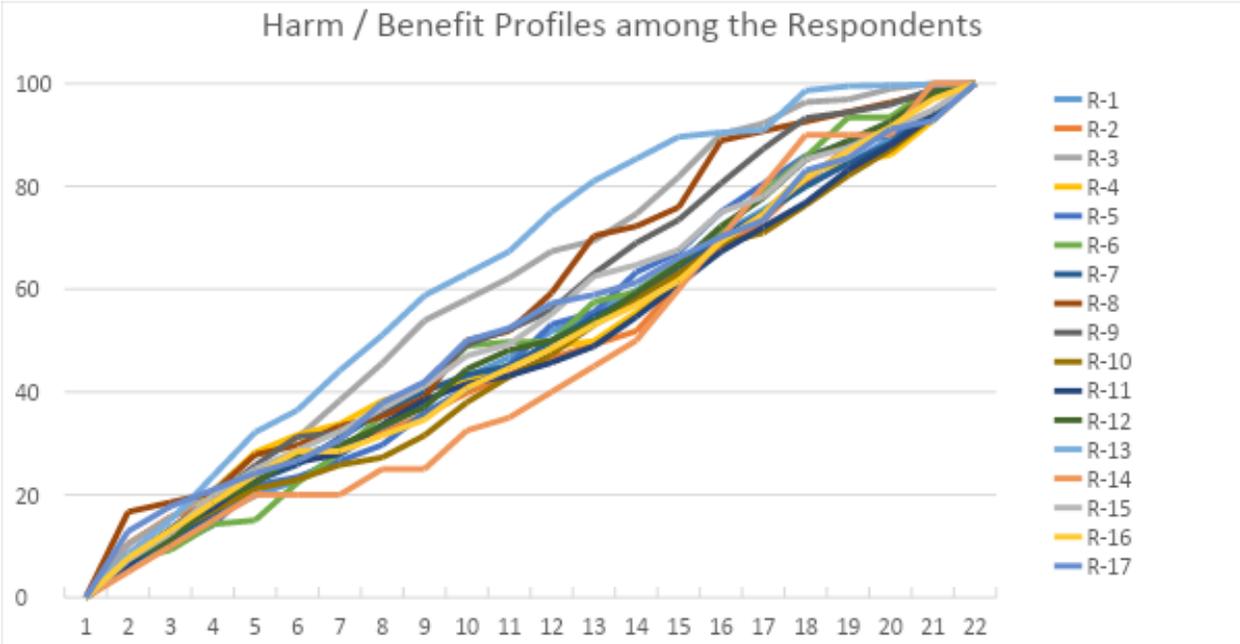

Figure S4: A CDF-like display of the harm / benefit assessments.

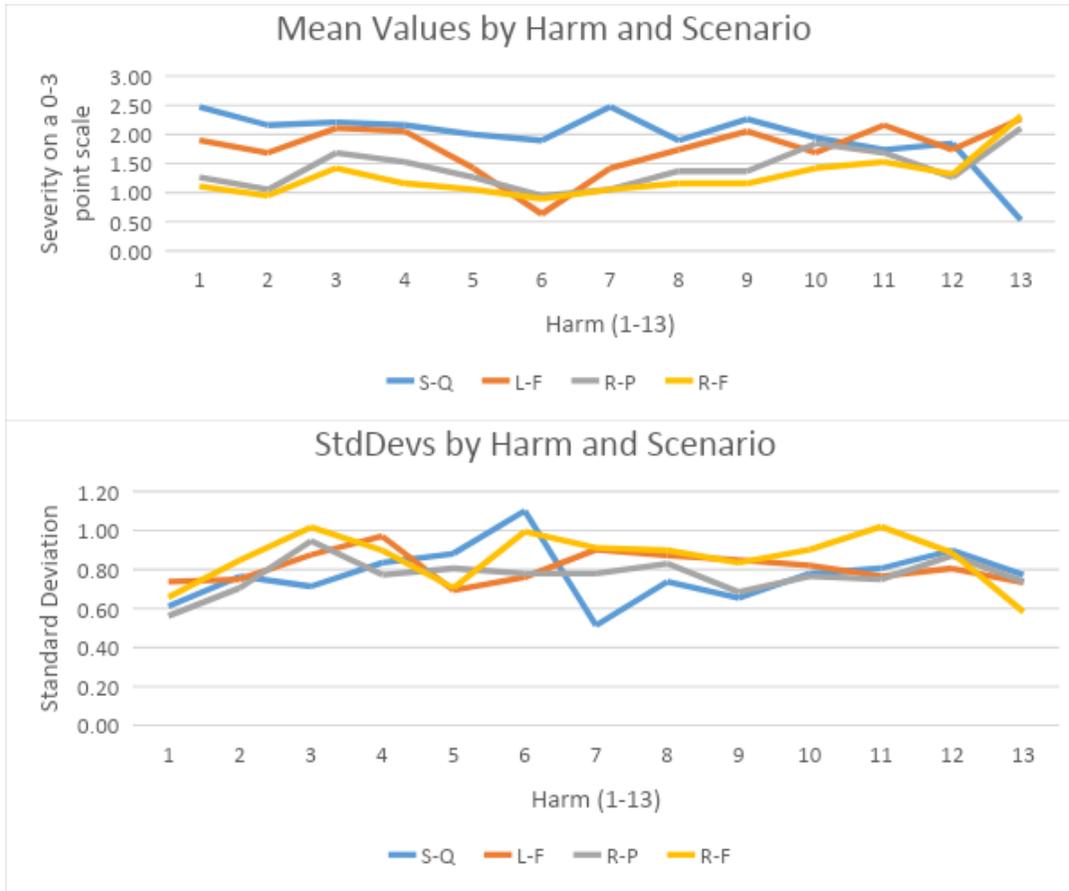

Figure S5: Means and standard deviations for 4-point assessments (0-3) by harm and scenario

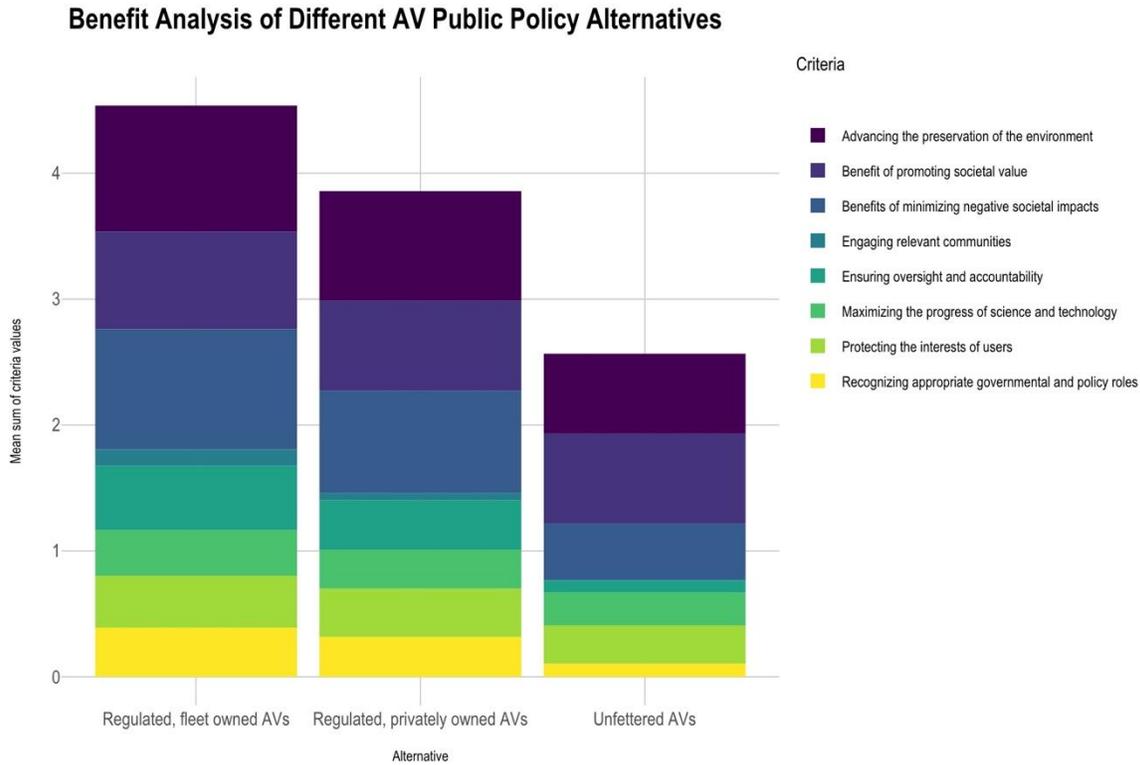

Figure S6: Mean values and standard deviations for 10-point assessments (0-9) by harm and scenario

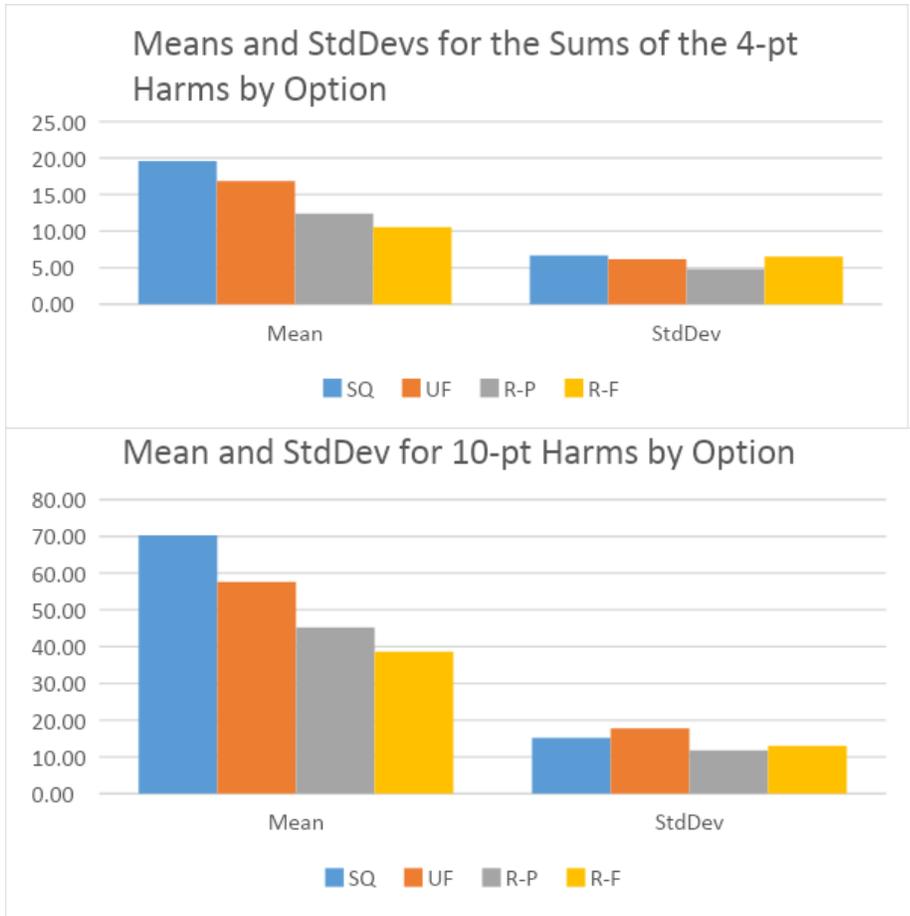

Figure S7: Means and standard deviations for the sums of the harms (by respondent) based on 4-point and 10-point ratings

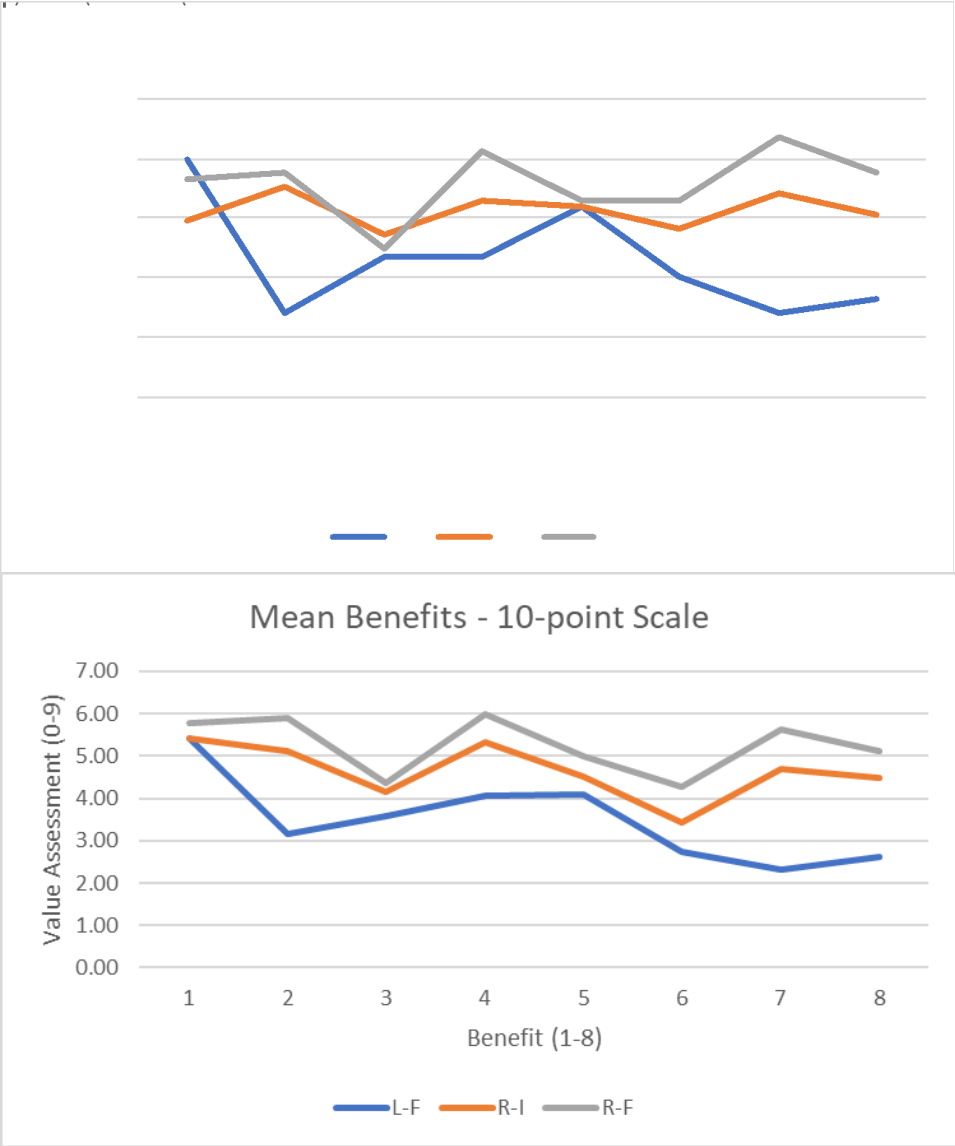

Figure S8: Average benefit value assessments for the 10-point scale